\documentclass[10pt]{article}
\usepackage{amsfonts}

\usepackage{graphics,epsfig}
\usepackage{graphicx}
\usepackage{epstopdf}
\newtheorem{theorem}{Theorem}
\newtheorem{definition}{Definition}

\begin{document}

\title{Compatibility and Separability for Classical \\
and Quantum Entanglement}
\author{Diederik Aerts, Christian de Ronde and Bart D'Hooghe \\
{\normalsize \itshape        Center Leo Apostel for Interdisciplinary
Studies (CLEA) }\\
{\normalsize \itshape         and Foundations of the Exact Sciences (FUND) }%
\\
{\normalsize \itshape
		Department of Mathematics, Vrije Universiteit Brussel }\\
{\normalsize \itshape         1160 Brussels, Belgium }\\
{\normalsize E-Mails: \textsf{diraerts@vub.ac.be, cderonde@vub.ac.be} }\\
{\normalsize \textsf{bdhooghe@vub.ac.be}}}
\date{}
\maketitle

\begin{abstract}
\noindent We study the concepts of compatibility and separability and their
implications for quantum and classical systems. These concepts are
illustrated on a macroscopic model for the singlet state of a quantum system
of two entangled spin 1/2 with a parameter reflecting indeterminism in the
measurement procedure. By varying this parameter we describe situations from
quantum, intermediate to classical and study which tests are compatible or
separated. We prove that for classical deterministic systems the concepts of
separability and compatibility coincide, but for quantum systems and
intermediate systems these concepts are generally different.
\end{abstract}

\section{Introduction}

Entanglement is one of the main features of quantum mechanics. The most
studied example of quantum entanglement is the one of the two spin 1/2
particles as presented by David Bohm \cite{bohm01}: more specifically in the
situation where the joint quantum entity of the two spin 1/2 particles is in
a singlet spin state. John Bell showed that quantum entanglement can be
experimentally tested by means of a set of inequalities, meanwhile called
Bell's inequalities \cite{bell01}. The `violation of Bell's inequalities'
indicates the presence of quantum entanglement. In \cite{aerts03} an example
of a situation consisting of macroscopic physical entities was proposed
where Bell's inequalities are violated. Later the example was elaborated to
yield a violation with exactly the same value $2\sqrt{2}$ as the one that
appears in the violation of Bell's inequalities by the spins in the singlet
spin state \cite{aerts10}. In the macroscopic example the considered
entities are two macroscopic particles, and the entanglement is expressed by
means of the presence of a rigid rod connecting the two particles. The rod
represents the non-local effect that appears with the violation of Bell's
inequalities. In \cite{aertsetal07} it is shown that not only the singlet
spin state, as considered in \cite{aerts03}, but all entangled states of
coupled spins can be represented by such a rod-like internal constraint. The
quantum probability is obtained in the model because of the introduction of
a hidden variable on the measurement apparatus in correspondence with the
spin model that was developed in \cite{aerts06,aerts07}.

The macroscopic model built in \cite{aerts03} violates Bell's inequalities
in a stronger way than the quantum spin example, indeed the violation value
for the macroscopic model is 4 while for the quantum example it is $2\sqrt{2}
$. We have proven in \cite{aertsetal01} that the quantum probability present
in the case of the quantum spin example, and also in the macroscopic example
of \cite{aerts03}, decreases the violation of Bell's inequalities. This is
the reason that a purely classical non-local situation, like the one
considered in \cite{aerts03}, violates Bell's inequalities in a stronger
way, with maximum possible violation value $4$.

We know that two measurements of which each one is performed on one of the
two spins of a quantum entangled spin couple are compatible measurements.
Indeed, self-adjoint operators $H_{1}\otimes I_{2}$ and $I_{1}\otimes H_{2}$
representing these measurements commute, which means that the corresponding
measurements are compatible. This means that a situation of quantum
entanglement can produce correlations between compatible measurements that
violate Bell's inequalities. This is not possible for a situation of
classical entanglements. Hence, if classical entanglement is present, at
least some of the measurements on the left and on the right will be
non-compatible measurements.

In the present article we investigate the aspect of compatibility and
separability as defined in section \ref{quantumclassicalentanglement} for
situations of classical and quantum entanglement. We show in section \ref%
{model} how increasing indeterminism, hence a change from classical
entanglement to quantum-like entanglement, increases compatibility for
measurements.

\section{Quantum and Classical Entanglement}

\label{quantumclassicalentanglement} If we want to study compatibility and
separability for macroscopic classical entities as well as for quantum
entities, we need to introduce a definition that is general enough to
capture all these situations. We do this in the framework of the
Geneva-Brussels approach to operational quantum axiomatics, where the basic
notions are the ones of tests, states and properties \cite%
{piron01,piron02,aerts01,aerts02,aerts04,aerts05,piron03,piron04}. Let us
consider the situation of a physical entity $S$. The entity $S$ is described
by a state space $\Sigma =\{p,q,\ldots \}$ and a set of tests $\mathcal{Q}%
=\{\alpha ,\beta ,\ldots \}$ which are defined by dichotomic experiments
with outcomes `yes' and `no' (see \cite{aerts02,aerts05} for the terminology
used). We denote $O(\alpha ,p)\subseteq \left\{ \hbox{`yes'},\hbox{`no'}%
\right\} $ the set of possible outcomes for the test $\alpha \in \mathcal{Q}$%
, the physical entity being in state $p\in \Sigma .$ Obviously the set of
possible outcomes of the test $\alpha $ is $O\left( \alpha \right) =\cup
_{p\in \Sigma }O(\alpha ,p)=\left\{ \hbox{`yes'},\hbox{`no'}\right\} .$ We
consider the situation of two physical entities $S_{1}$ and $S_{2}$, with
state spaces $\Sigma _{1}$ and $\Sigma _{2}$, and sets of tests $\mathcal{Q}%
_{{1}}$ and $\mathcal{Q}_{2}$. The two entities $S_{1}$ and $S_{2}$ form a
compound entity $S$, with state space $\Sigma $, and set of tests $\mathcal{Q%
}$. Suppose that the compound entity $S$ is in state $p\in \Sigma $, and
consider two tests $\alpha _{1}\in \mathcal{Q}_{1}$ and $\alpha _{2}\in 
\mathcal{Q}_{2}$. If tests $\alpha _{1}$ and $\alpha _{2}$ can be performed
together on the compound entity $S$ we denote this joint test $E\left(
\alpha _{1},\alpha _{2}\right) $. It will give rise to couples of outcomes $%
\left( x_{1},x_{2}\right) \in O\left( E\left( \alpha _{1},\alpha _{2}\right)
\right) $, such that $x_{1}\in O(\alpha _{1})$ and $x_{2}\in O(\alpha _{2})$%
. Let us denote the probability that an experiment $\alpha $ yields an
outcome $x_{\alpha }^{i}\in O\left( \alpha \right) $ for the entity $S$
being in the initial state $p\in \Sigma $ by $P(\alpha ,x_{\alpha }^{i}\mid
p).$

\subsection{Compatibility}

Suppose that one considers an experiment $E$ with four possible outcomes,
let's call them $x_{1},x_{2},x_{3}$ and $x_{4}$. Suppose now that we
consider two sub-experiments of this experiment $E$, by identifying some of
the outcomes. Hence $\alpha _{1}$ is the experiment that consists in
performing $E$, and giving outcome `yes' if we get $x_{1}$ or $x_{2}$ for $E$%
, and `no' if we get $x_{3}$ or $x_{4}$ for $E$. The experiment $\alpha _{2}$
consists in performing $E$, where we give `yes' if we get outcomes $x_{1}$
or $x_{3}$ and `no' if we get outcomes $x_{2}$ or $x_{4}$. Since both
subexperiments $\alpha _{1}$ and $\alpha _{2}$ consist in performing the big
experiment $E$, we also perform $\alpha _{1}$ and $\alpha _{2}$ together
whenever we perform $E$. Let us see how these two sub experiments are
related towards each other. Suppose $x_{2}$ is a possible outcome for $E,$
hence `yes, no' is a possible outcome for experiment $E.$ Then (by
definition) `yes' is possible for $\alpha _{1}$ and `no' is possible for $%
\alpha _{2}$. This is also the case for the other `yes' and `no'
combinations, which shows that $\alpha _{1}$ and $\alpha _{2}$ are
compatible as defined in \cite{aerts01},\cite{aerts11}. Hence sub
experiments of a big experiment defined in this straightforward way are
always compatible experiments. Compatibility expresses the situation where
the sub experiments behave as if they could be substituted by one big
experiment, after identification of certain outcomes. In terms of
probabilities of experiments, one can express the basis idea of
compatibility, namely compatible experiments can be substituted by one big
experiment such that they are recovered after identification of certain
outcomes, by following conditions:

\begin{definition}[Compatibility]
Tests $\alpha _{1}$ and $\alpha _{2}$ are `compatible' tests with respect to
a state $p\in \Sigma $ iff 
\begin{eqnarray}
P(\alpha _{1},\hbox{`yes'}\mid p) &=&P(E,x_{1}\mid p)+P(E,x_{2}\mid p)
\label{comp01} \\
P(\alpha _{1},\hbox{`no'}\mid p) &=&P(E,x_{3}\mid p)+P(E,x_{4}\mid p)
\label{comp02} \\
P(\alpha _{2},\hbox{`yes'}\mid p) &=&P(E,x_{1}\mid p)+P(E,x_{3}\mid p)
\label{comp03} \\
P(\alpha _{2},\hbox{`no'}\mid p) &=&P(E,x_{2}\mid p)+P(E,x_{4}\mid p)
\label{comp04}
\end{eqnarray}
\end{definition}

In case of compatibility, when for example $P(E,x_{1}\mid p)$ is different
from zero, hence (yes,yes) is a possible outcome, then $P(\alpha _{1},%
\hbox{`yes'}\mid p)$ and $P(\alpha _{2},\hbox{`yes'}\mid p)$ are different
from zero (because probabilities cannot be smaller than zero), hence `yes'
is a possible outcome for $\alpha _{1}$ and `yes' is a possible outcome for $%
\alpha _{2}$. Grasping the concept of compatibility in probabilistic terms
instead of merely `yes-no' statements as in \cite{aerts01},\cite{aerts11}
allows for a more detailed and realistic classification of the experiments,
in the sense that performing real experiments leads to probabilities rather
than yes-no statements.

\subsection{Separability}

Suppose that we start now with two experiments $\alpha _{1}$ and $\alpha _{2}
$, such that whatever one does with $\alpha _{1}$ does not influence what
one does with $\alpha _{2}$ (the idea of separated of Einstein Podolsky
Rosen, \cite{EPR}), then, just from pure logic, follows that whenever $%
\alpha _{1}$ has a possible outcome `yes' and for example $\alpha _{2}$ has
a possible outcome `no', then the joint experiments has as possible outcome
the couple `yes, no', because there is no influence. That is the basis of
separability.

\begin{definition}[Separability]
Two tests $\alpha _{1}$ and $\alpha _{2}$ are called `separated' with
respect to a state $p\in \Sigma $ iff 
\begin{eqnarray}
P(E,x_{1}\mid p) &=&P(\alpha _{1},\hbox{`yes'}\mid p)P(\alpha _{2},%
\hbox{`yes'}\mid p)  \label{sep01} \\
P(E,x_{2}\mid p) &=&P(\alpha _{1},\hbox{`yes'}\mid p)P(\alpha _{2},%
\hbox{`no'}\mid p)  \label{sep02} \\
P(E,x_{3}\mid p) &=&P(\alpha _{1},\hbox{`no'}\mid p)P(\alpha _{2},%
\hbox{`yes'}\mid p)  \label{sep03} \\
P(E,x_{4}\mid p) &=&P(\alpha _{1},\hbox{`no'}\mid p)P(\alpha _{2},\hbox{`no'}%
\mid p)  \label{sep04}
\end{eqnarray}
\end{definition}

\begin{definition}[Separated Tests]
Two tests $\alpha _{1}$ and $\alpha _{2}$ are `separated' iff $\forall p\in
\Sigma :\alpha _{1}$ and $\alpha _{2}$ are separated with respect to $p.$
\end{definition}

This means that if $x_{1}\in O(\alpha _{1},p)$ and $x_{2}\in O(\alpha
_{2},p) $ are possible outcomes for the tests $\alpha _{1}$ and $\alpha _{2}$
on the system $S$ in state $p$ then also the combined outcome $\left(
x_{1},x_{2}\right) $ is a possible outcome for the joint test $E\left(
\alpha _{1},\alpha _{2}\right) $. Suppose that for a specific state the
experiment $E$ only has two possible outcomes, namely $x_{2}$ and $x_{3}$,
then for this state the just constructed experiments $\alpha _{1}$ and $%
\alpha _{2}$ are not separated. Indeed if $E$ gives the outcome $x_{2}$ we
get `yes' for $\alpha _{1}$, and if $E$ gets the outcome $x_{3}$ we get `no'
for $\alpha _{1}$, which means that `yes' and `no' are possible outcomes for 
$\alpha _{1}$. In an analogous way it is easy to see that `yes' and `no' are
possible outcomes for $\alpha _{2}$. But `yes, yes' is not a possible
outcome for $E$, which proves that $\alpha _{1}$ and $\alpha _{2}$ are not
separated. In the example of `always compatible experiments' it is clear
that these experiments are not necessarily separated, since in this case
they are `always' executed together, which means that they are highly
dependent, and not at all independent.

Let us notice that separability implies compatibility. Indeed, putting
expressions (\ref{sep01}-\ref{sep04}) into (\ref{comp01}-\ref{comp04}) one
can easily verify that compatibility holds. A necessary and also sufficient
condition for two compatible experiments to be separated is the following: 
\begin{equation}
P(E,x_{1}\mid p)P(E,x_{4}\mid p)=P(E,x_{2}\mid p)P(E,x_{3}\mid p)
\label{CompatibleToSeparability}
\end{equation}
Clearly this is a necessary condition for separability, and after some
elementary calculations one can verify that it is also a sufficient
condition for two compatible experiments to be separated.

\subsection{Classicality}

\begin{definition}[Classicality]
A test $\alpha \in \mathcal{Q}$ is called a `classical test' iff $\forall
p\in \Sigma ,\exists x\left( p\right) \in O\left( \alpha \right) :$ $(\alpha
,p)=\left\{ x\left( p\right) \right\} .\label{defclassical}$
\end{definition}

Hence for each state $p$ a classical test $\alpha $ yields a specific
outcome $x\left( p\right) $ with certainty. Similarly a joint test $E\left(
\alpha _{1},\alpha _{2}\right) $ is classical iff $\forall p\in \Sigma
,\exists \left( x_{1},x_{2}\right) \left( p\right) \in O\left( E\left(
\alpha _{1},\alpha _{2}\right) \right) :$ $O(E\left( \alpha _{1},\alpha
_{2}\right) ,p)=\left\{ \left( x_{1},x_{2}\right) \left( p\right) \right\} .$

\begin{definition}[Classical Entity]
An entity is called a classical entity iff all its tests are classical.
\end{definition}

It is not possible to give an example of a classical entity with tests that
are compatible but not separated. This is because for classical entities the
two concepts are equivalent.

\begin{theorem}
If $\alpha _{1},\alpha _{2}$ are classical tests then $\alpha _{1}$ and $%
\alpha _{2}$ are compatible with respect to a state $p$ iff $\alpha _{1}$
and $\alpha _{2}$ are separated with respect to this state $p$.\label%
{classicaltheorem}
\end{theorem}

\noindent \textbf{Proof}: Suppose $\alpha _{1}$ and $\alpha _{2}$ are
classical tests that are compatible with respect to state $p$. Suppose `yes'
is a possible answer for $\alpha _{1}$ and `no' is a possible answer for $%
\alpha _{2}$ in state $p$. Since both $\alpha _{1}$ and $\alpha _{2}$ are
classical, these are the only respective outcomes for $\alpha _{1}$ and $%
\alpha _{2}$ such that $P(\alpha _{1},\hbox{`yes'}\mid p)=1=P(\alpha _{2},%
\hbox{`no'}\mid p).$ Consequently, $P(\alpha _{1},\hbox{`no'}\mid
p)=P(\alpha _{2},\hbox{`yes'}\mid p)=0.$ Because of compatibility, $%
P(E,x_{1}\mid p)=P(E,x_{3}\mid p)=P(E,x_{4}\mid p)=0,$ such that $%
P(E,x_{2}\mid p)=1.$ Hence the requirement \ref{CompatibleToSeparability} is
satisfied and tests $\alpha _{1}$ and $\alpha _{2}$ are separated tests.%
\mbox{} \hfill $\Box $

\noindent Hence for classical entities these concepts coincide,
such that compatible classical tests are always separated. Also notice that
classicality of the joint experiment follows from the compatibility of the
classical subtests. That a joint test constructed from two classical
subtests is not necessarily a classical test too, can be seen from following
example. Let us consider a slightly modified version of the example given in 
\cite{aerts03,aertsetal01}. The entity consists of two vessels connected by
a tube and contains 20 liters of transparent water. Tests $\alpha _{i}$ on
the left, respectively right vessel are defined by taking a siphon and
collecting water in a reference vessel. If we consider the test $\alpha _{i}$
defined as `can we take more than 10 liters of water from the vessel $i$'
then test $\alpha _{1}$ performed on the left vessel yields `yes' with
certainty. If on the other hand we would chose to perform the test $\alpha
_{2}$ on the right vessel then we would obtain `yes' with certainty. If we
define the joint test $E\left( \alpha _{1},\alpha _{2}\right) $ by
performing both tests $\alpha _{1}$ and $\alpha _{2}$ at the same time, we
can never find the answer (`yes',`yes') showing that tests $\alpha _{1}$ and 
$\alpha _{2}$ are not compatible and hence not separated. An example of a
joint test of two classical subtests which is not classical can be
constructed as follows. Let us consider the test $\beta _{2}$ on the right
vessel, defined as follows: `take a random amount of water (between 0,5 and
19,5 liter) out the right vessel with the fastest siphon in the universe and
see whether the water is transparent or not'. Because of the initial
assumption that the vessels are filled with transparent water, $\beta _{2}$
always yields the answer `yes'. Again, if we would perform test $\alpha _{1}$
we find the answer `yes' with certainty such that $P(\alpha _{1},\hbox{`yes'}%
\mid p)=1=P(\beta _{2},\hbox{`yes'}\mid p).$ On the other hand, if the joint
test $E\left( \alpha _{1},\beta _{2}\right) $ is performed by making both
tests at the same time, then in some cases we obtain the outcome $\left( %
\hbox{`yes'},\hbox{`yes'}\right) $ in the case the random amount of water
collected by test $\beta _{2}$ is less than 10 liters of water, and $\left( %
\hbox{`no'},\hbox{`yes'}\right) $ in the other case since test $\beta _{2}$
uses the fastest siphon and collects all the water it needs to check for
transparency before test $\alpha _{1}$ is able to collect more than 10
liters of water. Hence the joint test is not classical, since $P(E,x_{1}\mid
p)=P(E,x_{3}\mid p)=1/2.$ Also, tests $\alpha _{1}$ and $\beta _{2}$ are not
compatible, since $P(E,x_{2}\mid p)=P(E,x_{4}\mid p)=0,$ such that $P(\alpha
_{1},\hbox{`yes'}\mid p)=1\neq 1/2=P(E,x_{1}\mid p)+P(E,x_{2}\mid p).$ This
example shows that compatibility of classical tests is a very powerful
condition, since it not only implies separability of the two tests, but also
classicality of the joint test.

For the example of the connected vessels of water it is clear that the
considered tests cannot be separated because of the clear physical
connection between the two vessels. For entangled quantum systems we
encounter non-separated tests in a natural way (actually it is not possible
to describe a compound system of two separated quantum systems within the
formalism of quantum theory\cite%
{aerts01,aerts02,aerts04,aertsvalckenborgh01,aertsvalckenborgh02,aertsvalckenborgh03}%
), but the origin of this quantum non-separability is still a matter of
debate for physicists and philosophers alike. To shed more light on this
issue, we discuss in the next section a macroscopic model of a compound
system of two entangled spin 1/2 in the singlet state, in which tests are
defined by a parameter reflecting randomness in the measurement such that we
obtain a continuous transition from a quantum system to a classical system,
and check what happens with separability and compatibility for these systems.

\section{The Model}

\label{model}

For the individual spin 1/2 entities of the coupled spin 1/2 entity we use a
sphere model representation developed within the hidden measurement approach
to quantum mechanics \cite%
{aerts06,aerts07,aerts08,aerts09,aertsaerts01,aertsetal02}, which is a
generalization of the Poincar\'{e} or Bloch representation, such that also
the measurements as well as a parameter for non-determinism are described 
\cite{aertsetal07}.

In this model, a spin 1/2 entity is represented by a point particle on the
surface of a 3-dimensional unit sphere, called Poincar\'{e} sphere, such
that its state ${|\psi \rangle }=\left( \cos {\frac{\theta }{2}}e^{\frac{%
-i\phi }{2}},\sin {\frac{\theta }{2}}e^{\frac{i\phi }{2}}\right) $ is
represented by the point $u(1,\theta ,\phi )=(\sin \theta \cos \phi ,\sin
\theta \sin \phi ,\cos \theta )$. All points of the Poincar\'{e} sphere
represent states of the spin, such that points $u(1,\theta ,\phi )$ on the
surface correspond to pure states, while interior points $u(r,\theta ,\phi )$
correspond to density states $D(r,\theta ,\phi ):$ 
\begin{equation}
D(r,\theta ,\phi )={\frac{1}{2}}\left( 
\begin{array}{cc}
1+r\cos \theta  & r\sin \theta e^{-i\phi } \\ 
r\sin \theta e^{i\phi } & 1-r\cos \theta 
\end{array}%
\right) 
\end{equation}%
In this expression $D(1,\theta ,\phi )={|\psi \rangle }{\langle \psi |}$ is
the usual density state representation of a pure state, while $D(0,\theta
,\phi )$ is the density matrix representing the center of the sphere. 
\begin{figure}[h]
\centerline {\includegraphics[width=5cm]{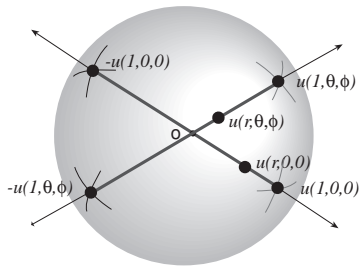}}
\caption{The Poincar\'{e} or Bloch sphere. Each point of the sphere
represents a state of the spin 1/2. We have indicated 6 states explicitly, $%
u(1,\protect\theta ,\protect\phi ),u(r,\protect\theta ,\protect\phi )$ and $%
-u(1,\protect\theta ,\protect\phi )$, which are three states in the spin
direction $(\protect\theta ,\protect\phi )$, and $u(1,0,0),u(r,0,0)$ and $%
-u(1,0,0)$, which are three states in the spin direction $(0,0)$. }
\label{spheremodel}
\end{figure}
Next to this, the sphere model allows a representation of measurements by
introducing an elastic band that is located along the measurement direction $%
u(\theta ,\phi )$ (we omit the radius $1$ since the measurement direction is
completely defined by angles $(\theta ,\phi )$). We introduce a test $\alpha
_{u(\theta ,\phi )}^{\epsilon }$ that is the following. We put a piece of
elastic of length $2\epsilon $ in the middle of the sphere and connect it
with `unbreakable' cords with the point $u(1,\theta ,\phi )$ on the surface
of the sphere and its diametrically opposite point $-u(1,\theta ,\phi )$
(Fig. \ref{spheremodel}). Suppose that the spin is in a state represented by
the point $u(r^{\prime },\theta ^{\prime },\phi ^{\prime })$. Once the
elastic is placed, the point particle falls orthogonally from its place $%
u(r^{\prime },\theta ^{\prime },\phi ^{\prime })$ onto the elastic, and
stays stuck to it in the point $u(1,\theta ,\phi )\cdot u(r^{\prime },\theta
^{\prime },\phi ^{\prime })$. Then the piece of elastic breaks with uniform
probability. If the elastic breaks in the interval $\left( -\epsilon
,u(1,\theta ,\phi )\cdot u(r^{\prime },\theta ^{\prime },\phi ^{\prime
})\right) $ the elastic pulls the particle to the point $u(1,\theta ,\phi )$
where it stays and the test $\alpha _{u(\theta ,\phi )}^{\epsilon }$ yields
answer `yes'. If on the other hand the elastic breaks in the interval $%
\left( u(1,\theta ,\phi )\cdot u(r^{\prime },\theta ^{\prime },\phi ^{\prime
}),\epsilon \right) $ the particle ends up in $-u(1,\theta ,\phi )$ and the
test $\alpha _{u(\theta ,\phi )}^{\epsilon }$ yields the answer `no'. 
\begin{figure}[h]
\centerline {\includegraphics[width=5cm]{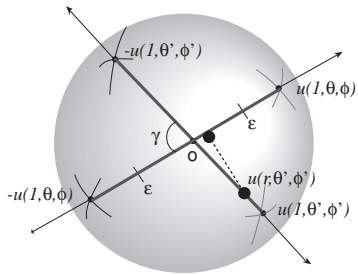}}
\caption{Between $-u(1,\protect\theta ,\protect\phi )$ and $u(1,\protect%
\theta ,\protect\phi )$ is a connection consisting of an elastic of length $2%
\protect\epsilon $ around $0$, and an unbreakable cord for the rest. The
point particle in $u(r^{\prime },\protect\theta ^{\prime },\protect\phi %
^{\prime })$ falls orthogonally on this connection. Then the elastic breaks
at random uniformly, and one of the two pieces of the connection drags the
point particle to one of the two points $-u(1,\protect\theta ,\protect\phi )$
or $u(1,\protect\theta ,\protect\phi )$, respectively with probabilities
that are the quantum spin 1/2 transition probabilities. }
\label{spheremodelmeasurement}
\end{figure}
If the elastic breaks exactly in the point $u(1,\theta ,\phi )\cdot
u(r^{\prime },\theta ^{\prime },\phi ^{\prime })$ we make the hypothesis
that the test $\alpha _{u(\theta ,\phi )}^{\epsilon }$ yields the outcome
`yes'. If $\epsilon =1,$ it can be proven that the sphere model reduces to a
model for the spin measurements on a spin 1/2 particle, since the
probabilities of the outcome `yes' (respectively `no') for the test $\alpha
_{u(\theta ,\phi )}^{1}$ coincide with the quantum probabilities for a spin
measurement with a Stern-Gerlach apparatus along the $(\theta ,\phi )$%
-direction (Fig. \ref{spheremodelmeasurement}) \cite%
{aerts06,aerts07,aerts08,aerts09,aertsetal07}. Also, the corresponding state
transitions are the same, namely a collapse from the initial state towards
an eigenstate of the observed outcome. If $\epsilon =0$, the model reduces
to a classical model: all tests $\alpha _{u(\theta ,\phi )}^{0}$ are
classical, in the sense of definition \ref{defclassical}. Remark that in
this classical limit ($\epsilon =0$) still a state transition occurs, but
from the probabilistic point of view measurements behave classical (i.e.
deterministic). For $0<\epsilon <1$ we find a situation in between quantum
and classical \cite%
{aertsetal06,aertsdurt01,aertsdurt02,aertsetal04,aertsetal05}. Assuming
uniform distribution of where the elastic breaks, probabilities are given by 
\begin{eqnarray*}
P(\alpha _{u(\theta ,\phi )}^{\epsilon },\hbox{`yes'}\mid p) &=&\frac{%
\epsilon +\cos \theta _{up}}{2\epsilon } \\
P(\alpha _{u(\theta ,\phi )}^{\epsilon },\hbox{`no'}\mid p) &=&\frac{%
\epsilon -\cos \theta _{up}}{2\epsilon }
\end{eqnarray*}%
for $-\epsilon <\cos \theta _{up}<\epsilon ,$ $P(\alpha _{u(\theta ,\phi
)}^{\epsilon },\hbox{`yes'}\mid p)=1$ for $\cos \theta _{up}\geq \epsilon ,$ 
$P(\alpha _{u(\theta ,\phi )}^{\epsilon },\hbox{`no'}\mid p)=1$ for $\cos
\theta _{up}\leq -\epsilon .$

\begin{figure}[h]
\centerline {\includegraphics[width=11cm]{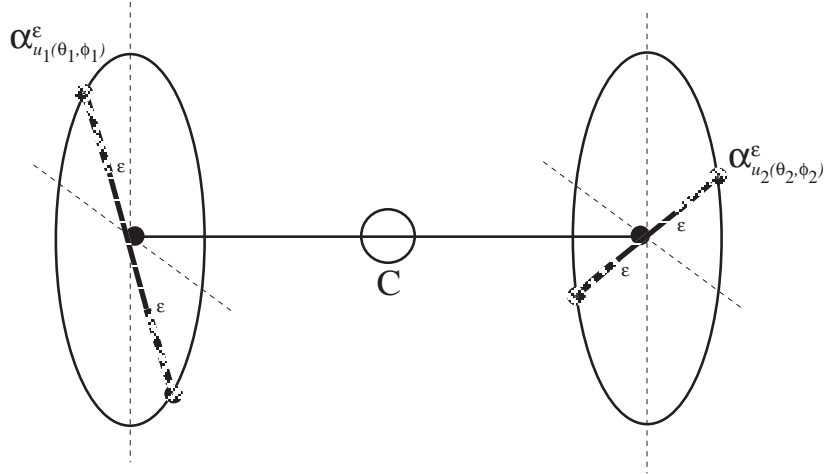}}
\caption{Symbolic representation of the singlet state in the model as two
dots in the centers of their respective spheres. Also shown is the
connecting rod and the measurement directions chosen at the left and right
location }
\label{macroentangledspins}
\end{figure}

The connected vessels of water entity is already an example of a macroscopic
system which violate Bell's inequalities \cite{aerts01}. However, the
expectation values it attains violate Bell's inequalities more than possible
for any quantum system of two entangled spin 1/2. In a way, this follows
from the fact that the connected vessels of water entity behaves `too
deterministic': if we allow indeterminism in the tests one can obtain
expectation values which still violate Bell's inequalities but within the
bounds of quantum probability \cite{aertsetal01}. In effect, in \cite%
{aerts09} a macroscopic system was considered which violates Bell's
inequalities in exactly the same way as a quantum system of two entangled
spin 1/2 in the singlet state. The model consists of two sphere models, such
that the point particles are connected by a rigid but extendable rod (Fig. %
\ref{macroentangledspins}). The singlet state is represented by the two
point particles being in the center of the respective Poincar\'{e} sphere,
which corresponds with a density state of the sphere model. 
A joint test $E\left( \alpha _{1},\alpha _{2}\right) $ on the compound
system is defined by performing first the test $\alpha _{1}$ and next the
test $\alpha _{2}$ as follows. First let us consider the quantum case with $%
\epsilon =1$ and show that the model indeed is a model for the singlet
state. When the first test $\alpha _{u(\theta _{1},\phi _{1})}^{1}$ is
performed, one of the two possible outcomes occurs with probability 1/2
since the point particle is in the middle of the sphere and the elastic
pulls the left point particle to the corresponding eigenstate. However,
because of the rod between the two particles, the other particle which
initially was in the center of the right sphere is drawn towards a point on
the surface of the sphere, namely the point diametrically opposite to the
eigenstate of the observed outcome of the first test. Only then the second
test is performed on the point particle, following the same description as
for a single sphere model. When the frequency of the coincidence counts is
calculated it turns out to be in exact accordance with the quantum
mechanical prediction for the spin measurements on a quantum system of two
entangled spin 1/2 in the singlet state \cite{aerts10,aertsetal01}.
Actually, because of symmetry in the initial singlet state it does not
matter which test is performed first (left or right), since the results
(i.e. probabilistic distribution over the set of outcomes and the
corresponding final states) are the same.

There are two key ingredients of this model that seem particularly important
in order to make it a model for the singlet state. First we have the rigid
rod, which shows the non-separable wholeness of the singlet coincidence
measurement (cfr. the role of the connecting tube between the vessels of
water) and which reflects the role of non-locality in quantum theory.
Secondly, we have the elastic that breaks which gives rise to the
probabilistic nature (cfr. the role of the siphon in the vessels of water)
of the outcomes which reflects the intrinsic indeterminism in quantum
experiments. Together these features establish the meaning of the violation
of Bell's inequalities as the `non-existence of local realism'. Let us now
explore the role of indeterminism for the concepts of separability and
compatibility by introducing the parameter $\epsilon $ for the compound
system in the singlet state in an analogous way as for a single spin 1/2
model (Fig \ref{macroentangledspins}).

We analyze for different values of the parameter $\epsilon $ which tests are
compatible or separated. Consider the case $\epsilon =1$, which means that
our model becomes a quantum model for two coupled spin 1/2 in the singlet
spin state. Consider the tests $\alpha _{u_{1}(\theta _{1},\phi _{1})}^{1}$
and $\alpha _{u_{2}(\theta _{2},\phi _{2})}^{1}$ for $\theta _{1}=\theta _{2}
$ and $\phi _{1}=\phi _{2}$, hence the case of two aligned spin directions.

Then the coupled sphere models are compatible but not separable: while for
each test $\alpha _{1}$ and $\alpha _{2}$ the result `yes' or `no' is
possible (with probability 1/2), the joint test $E\left( \alpha _{1},\alpha
_{2}\right) $ only has possible outcomes (`yes',`no') or (`no',`yes') for
the compound system in the singlet state, each with probability 1/2. Hence $%
P(E,x_{2}\mid p)=P(E,x_{3}\mid p)=1/2$ and the two tests are compatible but
not separable: $P(E,x_{1}\mid p)=0\neq 1/4=P(\alpha _{1},\hbox{`yes'}\mid
p)P(\alpha _{2},\hbox{`yes'}\mid p).$ The same situation occurs if we
consider the tests $\alpha _{u_{1}(\theta _{1},\phi _{1})}^{1}$ and $\alpha
_{u_{2}(\theta _{2},\phi _{2})}^{1}$, with $\theta _{1}=\pi -\theta _{2}$
and $\phi _{1}=\pi +\phi _{2}$ (i.e. $u_{1}$ and $u_{2}$ are opposite
directions): for both tests the result `yes' or `no' are possible outcomes
(with probability 1/2), while the joint test only has possible outcomes
(`yes',`yes') or (`no',`no') for the compound system in the singlet state.
Hence also these tests are compatible but not separable. We remark that if
we consider tests $\alpha _{u_{1}(\theta _{1},\phi _{1})}^{1}$ and $\alpha
_{u_{2}(\theta _{2},\phi _{2})}^{1}$ along different measurement directions $%
(\theta _{1},\phi _{1})$ and $(\theta _{2},\phi _{2})$, we find that these
tests are always compatible:%
\begin{eqnarray*}
P(E,x_{1}\mid p) &=&\frac{1}{2}\sin ^{2}\left( \frac{\theta _{uv}}{2}\right) 
\\
P(E,x_{2}\mid p) &=&\frac{1}{2}\cos ^{2}\left( \frac{\theta _{uv}}{2}\right) 
\\
P(E,x_{3}\mid p) &=&\frac{1}{2}\cos ^{2}\left( \frac{\theta _{uv}}{2}\right) 
\\
P(E,x_{4}\mid p) &=&\frac{1}{2}\sin ^{2}\left( \frac{\theta _{uv}}{2}\right) 
\end{eqnarray*}%
but that they are only separable if $\theta _{u_{1}u_{2}}=\frac{\pi }{2}$ in
which case the second measurement yields the same probability distribution
after performing the first experiment as for a single measurement, and it is
as if the subtests could be performed independently of each other.

Let us now consider the case $0<\epsilon <1$ and determine which tests are
compatible, respectively separable. Let us first notice that for the system
prepared in the singlet state we have $P\left( \alpha _{i},\hbox{`yes'}%
\right) =P\left( \alpha _{i},\hbox{`no'}\right) =1/2.$ Following the
description of the compound tests $E\left( \alpha _{u_{1}(\theta _{1},\phi
_{1})}^{\epsilon },\alpha _{u_{2}(\theta _{2},\phi _{2})}^{\epsilon }\right) 
$ for the system prepared in the singlet state, the point particle in the
second sphere will be pulled by means of the rod to the point $%
-u_{1}(1,\theta _{1},\phi _{1})$ on the sphere in the event of an answer
`yes' in the first test, and to the point $u_{1}(1,\theta _{1},\phi _{1})$
in the event of an outcome `no'. Hence if in the first subtest of the joint
experiment $E\left( \alpha _{u_{1}(\theta _{1},\phi _{1})}^{\epsilon
},\alpha _{u_{2}(\theta _{2},\phi _{2})}^{\epsilon }\right) $ the outcome
`yes' was found, then outcome `yes' occurs in the second measurement with
probability 
\begin{equation}
P(\alpha _{2},\hbox{`yes'}\mid -u_{1}(1,\theta _{1},\phi _{1}))=\frac{%
\epsilon +\cos \theta _{-u_{1}u_{2}}}{2\epsilon }=\frac{\epsilon -\cos
\theta _{u_{1}u_{2}}}{2\epsilon }
\end{equation}%
if $-\epsilon <\cos \theta _{u_{1}u_{2}}<\epsilon $, such that $%
P(E,x_{1}\mid p)=\frac{1}{2}\frac{\epsilon -\cos \theta _{u_{1}u_{2}}}{%
2\epsilon }.$ If $\cos \theta _{u_{1}u_{2}}\geq \epsilon $ we obtain $%
P(E,x_{1}\mid p)=0,$ and if $\cos \theta _{u_{1}u_{2}}\leq -\epsilon $ then $%
P(E,x_{1}\mid p)=1/2.$ Similarly, we find for $-\epsilon \leq \cos \theta
_{u_{1}u_{2}}\leq \epsilon $%
\begin{eqnarray*}
P(E,x_{1}\mid p) &=&\frac{1}{2}\frac{\epsilon -\cos \theta _{u_{1}u_{2}}}{%
2\epsilon } \\
P(E,x_{2}\mid p) &=&\frac{1}{2}\frac{\epsilon +\cos \theta _{u_{1}u_{2}}}{%
2\epsilon } \\
P(E,x_{3}\mid p) &=&\frac{1}{2}\frac{\epsilon +\cos \theta _{u_{1}u_{2}}}{%
2\epsilon } \\
P(E,x_{4}\mid p) &=&\frac{1}{2}\frac{\epsilon -\cos \theta _{u_{1}u_{2}}}{%
2\epsilon }
\end{eqnarray*}%
Clearly, the two experiments $\alpha _{u_{1}(\theta _{1},\phi
_{1})}^{\epsilon }$ and $\alpha _{u_{2}(\theta _{2},\phi _{2})}^{\epsilon }$
are always compatible but only separated if $\cos \theta _{u_{1}u_{2}}=0,$
i.e. $\theta _{u_{1}u_{2}}=\frac{\pi }{2}.$

Finally, let us see what happens in the classical situation $\epsilon =0$
for $\alpha _{u_{1}(\theta _{1},\phi _{1})}^{0},\alpha _{u_{2}(\theta
_{2},\phi _{2})}^{0}$. The first test $\alpha _{u_{1}(\theta _{1},\phi
_{1})}^{0}$ is deterministic since it yields answer `yes' with certainty,
such that the second particle is pulled towards the state $-u_{1}(1,\theta
_{1},\phi _{1})$. If $u_{2}(1,\theta _{2},\phi _{2})\cdot u_{1}(1,\theta
_{1},\phi _{1})>0$ the second test $\alpha _{u_{2}(\theta _{2},\phi
_{2})}^{0}$ yields answer `no' with certainty. If $u_{2}(1,\theta _{2},\phi
_{2})\cdot u_{1}(1,\theta _{1},\phi _{1})\leq 0$ the second test will yield
the outcome `yes' with certainty. Hence depending on the angle between the
measurement directions $u_{2}(1,\theta _{2},\phi _{2})$ and $u_{1}(1,\theta
_{1},\phi _{1})$ we obtain that the set of possible outcomes for the joint
test $E\left( \alpha _{u_{1}(\theta _{1},\phi _{1})}^{0},\alpha
_{u_{2}(\theta _{2},\phi _{2})}^{0}\right) =\left\{ \left( \hbox{`yes'},%
\hbox{`no'}\right) \right\} $ if $u_{2}(1,\theta _{2},\phi _{2})\cdot
u_{1}(1,\theta _{1},\phi _{1})>0$ and $E\left( \alpha _{u_{1}(\theta
_{1},\phi _{1})}^{0},\alpha _{u_{2}(\theta _{2},\phi _{2})}^{0}\right)
=\left\{ \left( \hbox{`yes'},\hbox{`yes'}\right) \right\} $ if $%
u_{2}(1,\theta _{2},\phi _{2})\cdot u_{1}(1,\theta _{1},\phi _{1})\leq 0.$
Since for $\epsilon =0$ both test $\alpha _{u_{1}(\theta _{1},\phi _{1})}^{0}
$ and $\alpha _{u_{2}(\theta _{2},\phi _{2})}^{0}$ yield positive outcome
with certainty: $P\left( \alpha _{u_{1}(\theta _{1},\phi _{1})}^{0},%
\hbox{`yes'}\right) =1=P\left( \alpha _{u_{2}(\theta _{2},\phi _{2})}^{0},%
\hbox{`yes'}\right) $, we can conclude that for $u_{2}(1,\theta _{2},\phi
_{2})\cdot u_{1}(1,\theta _{1},\phi _{1})\leq 0$ the two tests are
compatible, and from theorem \ref{classicaltheorem} (or simple verification)
follows that they are also separable. For $u_{2}(1,\theta _{2},\phi
_{2})\cdot u_{1}(1,\theta _{1},\phi _{1})>0$ we have $P(E,x_{2}\mid
p)=1=P(\alpha _{u_{1}(\theta _{1},\phi _{1})}^{0},\hbox{`yes'}\mid
p)=P(\alpha _{u_{2}(\theta _{2},\phi _{2})}^{0},\hbox{`yes'}\mid p)$. Hence
tests along these directions are neither compatible nor separable, which is
in agreement with theorem \ref{classicaltheorem} which states that for
classical tests the concepts of compatibility and separability coincide.

\section{Conclusions}

We have studied a macroscopic model for the singlet state of a quantum
system of two spin 1/2 consisting of two sphere models connected by a rigid
but extendable rod. By considering a parameter representing indeterminism in
the measurement procedure, we have described a continuous transition from a
quantum to a classical system and shown which tests are compatible or
separable. We have proven that for classical, i.e. deterministic, systems
the concepts of separability and compatibility coincide, and that the joint
test of two compatible classical subtests is classical. As an illustration
we considered an example of a nonclassical joint test of two classical but
noncompatible subtests. For the compound system in the quantum limit tests
along non orthogonal measurement directions are always compatible but not
separable. Therefore, these tests and a fortiori the subsystems are not
separated, which is in agreement with previous results stating that
separated quantum entities cannot be described in quantum theory \cite%
{aerts01,aerts02,aerts04,aertsvalckenborgh01,aertsvalckenborgh02,aertsvalckenborgh03}%
. For intermediate values of the parameter we found that all tests are
compatible, but they are only separable if $\theta _{u_{1}u_{2}}=\frac{\pi }{%
2}.$ In the classical limit $\epsilon =0$ tests $\alpha _{u_{1}(\theta
_{1},\phi _{1})}^{0}$ and $\alpha _{u_{2}(\theta _{2},\phi _{2})}^{0}$ are
compatible and therefore also separable if $u_{2}(1,\theta _{2},\phi
_{2})\cdot u_{1}(1,\theta _{1},\phi _{1})\leq 0$. In these cases, the two
subsystems behave as if they are really separated and their is no rigid rod
connecting the two point particles. If $u_{2}(1,\theta _{2},\phi _{2})\cdot
u_{1}(1,\theta _{1},\phi _{1})>0$ tests $\alpha _{u_{1}(\theta _{1},\phi
_{1})}^{0}$ and $\alpha _{u_{2}(\theta _{2},\phi _{2})}^{0}$ are not
compatible and in accordance with theorem \ref{classicaltheorem} also not
separated, the role of the rigid rod cannot be neglected.

\end{document}